\documentclass[preprintnumbers,amsmath,amssymb]{revtex4}
\usepackage{graphicx}
\usepackage{dcolumn}
\usepackage{bm}
\begin{document}
\def\oppropto{\mathop{\propto}} 
\def\opsimeq{\mathop{\simeq}}
\def\opoverderline{\mathop{\overline}}
\def\operarrow{\mathop{\longrightarrow}}
\def\opsim{\mathop{\sim}} 

\title{Two-dimensional wetting with binary disorder: \\
a numerical study of the loop statistics }
\author{Thomas Garel and C\'ecile Monthus}
 \affiliation{Service de Physique Th\'{e}orique, CEA/DSM/SPhT\\
Unit\'e de recherche associ\'ee au CNRS\\
91191 Gif-sur-Yvette cedex, France}


\begin{abstract}
We numerically study the wetting (adsorption) transition 
of a polymer chain on a disordered substrate in $1+1$ dimension.
Following the Poland-Scheraga model of DNA denaturation, we use a
Fixman-Freire scheme for the entropy of loops. This allows us to consider
chain lengths of order $N \sim 10^5 $ to $10^6$, with $10^4$
disorder realizations. Our study is based on  the
statistics of  loops between two contacts with the
substrate, from which we define Binder-like parameters: their
crossings for various sizes $N$ allow a precise
determination of the critical temperature, and their finite size
properties yields a crossover exponent $\phi=1/(2-\alpha) \simeq 0.5$.
We then analyse at criticality  the distribution of loop length $l$
in both regimes $l \sim O(N)$ and $1 \ll l \ll N$, as well as the
finite-size properties
of  the  contact density and energy. Our conclusion is that the critical
exponents for the thermodynamics are the same as those of the pure
case, except for strong logarithmic corrections to scaling. The
presence of these logarithmic corrections in the thermodynamics is
related to a disorder-dependent logarithmic singularity that appears in
the critical loop distribution in the rescaled variable $\lambda=l/N$
as $\lambda \to 1$.

\bigskip


\end{abstract}
\maketitle

\section{Introduction}
The effect of disorder on the wetting transition in dimension $1+1$
has attracted a lot of interest in the last twenty years and remains a
rather controversial issue \cite{FLNO,Der_Hak_Van}.
The wetting model that we consider here is defined as
follows. The (impenetrable) substrate is located at $z=0$. The
polymer chain has $N$ monomers, and the position $z_{\alpha}$ of monomer
$(\alpha)$ satisfies $z_{\alpha} \ge 0$, with $z_1=z_N=0$. The
partition function of the model reads
\begin{equation}
\label{partit1}
Z=\sum_{(RW)} e^{-\beta H}
\end{equation}
where $H=\sum_{\alpha=1}^{N} \varepsilon_{\alpha}\
\delta_{z_{\alpha},0}$. In equation (\ref{partit1}), the sum runs over
all random walks (RW) with $\vert z_{\alpha+1}-z_{\alpha} \vert =\pm
1$ and $\beta =\frac{1}{k_B T}$ is the inverse 
temperature. The contact energies ($\varepsilon_{\alpha}$) are
independent quenched random variables. We study here the binary distribution
($\varepsilon=0$ with probability $p$ and $\varepsilon=\varepsilon_{0}$ 
with probability $1-p$).

On the analytical side, efforts have focused on the
small disorder limit \cite{FLNO,Der_Hak_Van}. Since the pure wetting
transition has a specific heat exponent $\alpha_{pure}=0$, the
disorder is marginal according to the Harris criterion \cite{Harris}.
Based on perturbative calculations, Ref \cite{FLNO} finds a marginally
irrelevant disorder (i) the 
quenched and annealed critical temperatures coincide
(ii) the quenched critical properties are the same as in the pure (or
annealed) case, up to subdominant logarithmic corrections. Other
studies have concluded that that the disorder is marginally relevant
\cite{Der_Hak_Van,Bhat_Muk,Ka_La} (i) the quenched and annealed
critical temperatures differ by a term which has an essential singularity
in the disorder strength \cite{Der_Hak_Van} (ii) the critical
behavior is governed by some non-trivial disordered fixed point. On
the numerical side, the same debate on the disorder 
relevance took place. Numerical studies for flat and
exponential disorder distributions \cite{FLNO}, or for
binary disorder distribution \cite{Cu_Hwa} have concluded that the
critical behavior was indistinguishable from the pure transition. On
the other hand, the numerical study of \cite{Der_Hak_Van} for binary
disorder pointed towards a negative specific heat exponent
($\alpha<0$). Finally, the study of Gaussian disorder \cite{Ta_Cha}
has been interpreted as an essential singularity in the specific heat,
that formally corresponds to an exponent $\alpha=-\infty $. 
This paper aims at clarifying the situation for the problem defined in
equation (\ref{partit1}), via the analysis of loop statistics
between two contacts with the substrate. 

\section{Poland-Scheraga model of the wetting transition}

\subsection{Model and observables}

The wetting model of eq. (\ref{partit1}) is equivalent to the Poland-Scheraga
description of DNA denaturation \cite{Pol_Scher,WB85}. The relation
between the two problems is made apparent if 
one interprets the coordinate $z$ as the relative coordinate of the two DNA
strands. We accordingly define a forward partition function
$Z_f(\alpha)$  for a chain of $\alpha$ monomers, with 
$z_1=z_{\alpha}=0$. From equation (\ref{partit1}) we get 
\begin{equation}
\label{forward}
Z_f(\alpha)=e^{-\beta \varepsilon _{\alpha}} \sum_{\alpha ^{\prime
}=1}^{\alpha -2}Z_f(\alpha ^{\prime }) 
\mathcal{N}(\alpha ^{\prime };\alpha ) 
\end{equation}
where $\mathcal{N}(\alpha ^{\prime };\alpha )$ is the the partition
function of a loop going from $\alpha ^{\prime }$ to $\alpha$. The
asymptotic expansion of $\mathcal{N}(\alpha ^{\prime };\alpha )$ is
given by \cite{PGG}
\begin{equation}
\label{asymp}
\mathcal{N}(\alpha ^{\prime };\alpha ) \simeq \sigma_0
 \ 2^{\alpha-\alpha ^{\prime}}\ f(\alpha-\alpha ^{\prime})
\end{equation}
where $\sigma_0$ is a constant and $f(x)=\frac{1}{x^c}$ is the
probability to return to the substrate after x steps. In our model,
$c=\frac{3}{2}$. Other values of $c$ are of interest in the DNA
denaturation problem \cite{Ka_Mu_Pe}. In a similar way, we define a
backward partition function $Z_b(\alpha)$, 
defined as the partition function of a chain of $N-\alpha$ monomers,
with $z_{\alpha}=z_N=0$, which satisfies
\begin{equation}
\label{backward}
Z_b(\alpha )=e^{-\beta \varepsilon _{\alpha}} \sum_{\alpha ^{\prime
}=\alpha+2}^{N}Z_b(\alpha ^{\prime }) 
\mathcal{N}(\alpha; \alpha ^{\prime }) 
\end{equation}
In these notations, the partition function $Z$ of equation
(\ref{partit1}) is given by $Z=Z_f(N)=Z_b(1)$, and the probability for 
monomer $\alpha$ to be adsorbed on the substrate is
\begin{equation}
\label{proba1}
p(\alpha)=\frac {Z_f(\alpha)Z_b(\alpha) e^{\beta
\varepsilon_{\alpha}}}{Z_f(N)}
\end{equation}
where the factor $e^{\beta \varepsilon_{\alpha}}$ in the numerator
avoids double counting of the contact energy at $\alpha$. The contact
density on the substrate (a quantity of primary importance in the DNA
context) is given by 
\begin{equation}
\theta_N(T) = \frac{1}{N} \sum_{i=1}^N p(\alpha)
\label{theta}
\end{equation}
In the pure case, $\theta_N(T)$ is proportional to the energy. Since
this is not true in the disordered case, we also consider the contact
energy 
\begin{equation}
e_N(T) = \frac{1}{N} \sum_{i=1}^N \varepsilon_{\alpha} \ p(\alpha)
\label{energy}
\end{equation}

We will also be interested in $P_{\rm
loop}(\alpha,\gamma)$, defined as the probability of having
a loop between monomers $\alpha$ and $\gamma$ on the substrate
\begin{equation}
\label{proba2}
P_{\rm loop}(\alpha,\gamma)=\frac
{Z_f(\alpha)\mathcal{N}(\alpha;\gamma)Z_b(\gamma)}{Z_f(N)} 
\end{equation}
\subsection{Numerical implementation}
The above equations, explained in more detail in \cite{Ga_Or}, show
that numerical calculations of 
the partition function $Z$  will require a CPU time of order $O(N^2)$. 
The Fixman-Freire method \cite{Fix_Fre} reduces this CPU time to
$O(N)$ by approximating the probability factor $f(x)$ of equation
(\ref{asymp}) by  
\begin{equation}
\label{FF}
f(x)=\frac{1 }{{x}^{3/2}} \simeq f_{FF}(x)=\sum_{i=1}^I a_i \
e^{-b_i x} 
\end{equation}
In equation (\ref{FF}) the number ${I}$ of couples $(a_{i},b_{i})$ depends
on the desired accuracy. The parameters $(a_{i},b_{i})$ are determined by a
set of non-linear equations.
This procedure has been tested on DNA chains of length up to $N=10^6$
base pairs \cite{Meltsim,Yer}, and the choice ${I}=15$ gives an accuracy   
better than $0.3\%$. We have adopted this value throughout this paper. 

Putting everything together, the model we have numerically
studied is defined by recursion equations
(\ref{forward},\ref{backward}) for the partition functions where the
loop partition function $\mathcal{N}(\alpha; \alpha 
^{\prime })$ has been replaced by its asymptotic expression
(\ref{asymp}), with the Fixman-Freire approximation (eq. \ref{FF})
for $f(x)$.

\section{Localization of the critical temperature}

\subsection{ Loop statistics and Binder-like parameters }

We now define the probability measure $M_N(l)$
for the loops existing in a sample of length $N$ as follows: for each
loop length $l$, we sum the loop probability $P_{\rm loop}
(\alpha,\alpha+l)$ (eq.(\ref{proba2})) over all possible origins
$(\alpha)$ 
\begin{eqnarray}
M_N(l) = \sum_{\alpha=1}^{N-l} P_{\rm loop} (\alpha,\alpha+l) 
\label{pnl}
\end{eqnarray}

The normalization of this measure over $l$ corresponds
to the averaged number of loops in a sample of size $N$,
or equivalently to the averaged number $N \theta_N (T)$ 
of contacts (\ref{theta}) with the substrate :
\begin{eqnarray}
M_N \equiv  \int dl M_N(l) =  N \theta_N  
\label{thetapl}
\end{eqnarray} 
This number is thus extensive $M_N \propto N$
 in the localized phase $(T<T_c)$, and remains finite as $N \to \infty$
in the delocalized phase ($T>T_c$).

The first moment of the loop measure $M_N(l)$
\begin{eqnarray}
<l>_N \equiv  \frac{ \int dl \ l M_N(l) }{ \int dl M_N(l)}  
\label{defm1}
\end{eqnarray} 
remains finite as $N \to \infty$ in the localized phase $(T<T_c)$,
whereas it diverges as $<l>_N \sim N$ in the delocalized phase ($T>T_c$).
 We thus
introduce the rescaled variable 
\begin{equation}
\lambda=\frac{l}{N}
\end{equation}
and the corresponding probability measure
${\cal M}_N(\lambda)$ for the loops occupying a finite fraction
$\lambda=l/N$ of the whole sample.
This measure ${\cal M}_N(\lambda)$ converges
respectively in the thermodynamic
limit towards $\delta(\lambda)$ in the localized phase $(T<T_c)$
and towards $\delta(\lambda-1)$ in the delocalized phase $(T>T_c)$.
 At the critical point $T=T_c$, ${\cal M}_N(\lambda)$
converges for large $N$ 
towards a stable measure ${\cal M}_c(\lambda)$
with support $\lambda \in \ ]0,1]$, in addition to
a singular part in $\delta (\lambda)$
whose weight represents the number of finite loops at criticality.

To locate the critical temperature,  it is thus convenient to
introduce the following Binder-like parameters \cite{Bind}
 \begin{eqnarray}
\label{Binder}
B_N(T)= \frac{<l^2>}{<l> N} \ \ \ ,  \ \ \  R_N(T)= \frac{<l^3><l>}{<l^2>^2} 
\label{defbtrue}
\end{eqnarray} 
where $<l>,\ <l^2>, \ <l^3>$ are the first moments of the measure
$M_N(l)$.
In the thermodynamic limit $N \to \infty$, the parameter
$B_{\infty}(T)$
jumps from $B_{\infty}(T<T_c)=0$ to $B_{\infty}(T>T_c)=1$.
For finite chain lengths ($N_1,N_2,...$), the ratios $B_{N_1}(T)$,
$B_{N_2}(T)$ .. cross at $T_c$ at some intermediate value $0<B(T_c)<1$
given by 
 \begin{equation}
\label{Bc}
B(T_c)= \frac{<\lambda^2>_c}{<\lambda>_c }
\end{equation} 
where $<\lambda^m>_c=\int d\lambda \ \lambda^m  \ {\cal M}_c(\lambda)$
are the moments of the critical loop measure ${\cal M}_c(\lambda)$.
The parameter $R_N(T)$ has a similar behavior with the following
crossing value 
 \begin{equation}
\label{Rc}
R(T_c)= \frac{<\lambda>_c \ <\lambda^3>_c}{<\lambda^2>_c^2 }
\end{equation}

Computerwise, the evaluation of the full loop distribution $l=1, ..,
N$ via eq.(\ref{pnl}) requires a time growing as $N^2$. To  keep
a computation time of order $O(N)$, we have chosen to sample the 
rescaled loop
measure ${\cal M}_N(\lambda)$
on a fixed number $k_{max}$ of values 
$\lambda_k=k/k_{max}$ with $k=1,2,..,k_{max}$.
 From this sampling of the loop distribution,
we define the reduced moments 
 \begin{eqnarray}
<\lambda^m>_{k_{max}}&&
=\sum_{k=1}^{k_{max}} \left( \frac{k}{k_{max}} \right)^m 
{\cal M}_N \left(\lambda_k= \frac{k}{k_{max}} \right) \\
\label{moments}
\end{eqnarray}
from which we build the modified Binder parameters
 \begin{eqnarray}
B_N^{(k_{max})}(T)= \frac{<\lambda^2>_{k_{max}}}{ \ <\lambda>_{k_{max}}} \ \ \ \ , 
\ \ \ R_N^{(k_{max})}(T)=
\frac{<\lambda^3>_{k_{max}}<\lambda>_{k_{max}}}{<\lambda^2>_{k_{max}}^2} 
\label{defb}
\end{eqnarray} 
whose properties are the same as the true Binder parameters
(\ref{defbtrue}) described above, except for the values of the crossing
points that now depend on $k_{max}$
 \begin{eqnarray}
B^{(k_{max})}(T_c)=
 \frac{<\lambda^2>_{c,k_{max}}}{ \ <\lambda>_{c,k_{max}}} \ \ \ \ , 
\ \ \ R^{(k_{max})}(T_c)=
\frac{<\lambda^3>_{c,k_{max}}<\lambda>_{c,k_{max}}}{<\lambda^2>_{c,k_{max}}^2} 
\label{crossing}
\end{eqnarray} 
where 
 \begin{equation}
<\lambda^m>_{c,k_{max}}
=\sum_{k=1}^{k_{max}} \left( \frac{k}{k_{max}} \right)^m 
{\cal M}_c(\lambda_k= \frac{k}{k_{max}} ) \\
\label{moments2}
\end{equation}

We now illustrate these notions on the pure case, before we turn to
the analysis of the disordered case.

\subsection{Loop statistics in the pure case}

In the $(\pm 1)$ random walk model (\ref{partit1}) with a pure substrate
($\varepsilon_{\alpha}=\varepsilon_0$), criticality
corresponds to the condition $e^{\beta_c \varepsilon_0}=2$
where the substrate is exactly reflexive \cite{Der_Hak_Van}.
The reflexive nature of the substrate at criticality
holds more generally for pure wetting models
from a functional RG analysis  \cite{Huse}.
This means that the partition function $Z_N$ with both ends fixed on the substrate is simply given by the number of random walks 
returning to the origin after $N$ steps
 \begin{equation}
Z_N^{pure}(T_c) \simeq \frac{ 2^N }{ \sqrt N }
\label{znpurtc}
\end{equation}
As a consequence, the critical loop measure (\ref{pnl}) reads
 \begin{equation}
M_N^{T_c}(l) \simeq \frac{\sqrt N }{ l^{3/2} } \int_1^{N-l}   \ \frac{d \alpha }{ {\sqrt \alpha} {\sqrt {N-l-\alpha } }}
 \simeq \frac{\sqrt N }{ l^{3/2} }
\label{critilpur}
\end{equation}
i.e. at criticality, there are $\sqrt N$ loops, whose lengths
are distributed with the random walk first return probability
 \begin{equation}
\rho(l) \sim \frac{ 1 }{ l^{3/2} }
\label{3sur2}
\end{equation}
In terms of the rescaled variable $\lambda=l/N$, the 
loop measure (\ref{critilpur}) becomes independent of the size $N$
 \begin{equation}
{\cal M}_c(\lambda)  \simeq \frac{ 1 }{ \lambda^{3/2} }
\label{critilambdapur}
\end{equation}
This means that at criticality, there are a finite number of loops
whose length $l$ represents a finite fraction of the size $N$ of the sample.
This property can be understood as follows : 
the L\'evy sum 
of $n$ independent random variables distributed with (\ref{3sur2})
scales as $l_1+l_2+ ... l_n \sim n^{2}$ : as a consequence, the number $n$
of loops scales with the size $N \sim l_1+l_2+ ... l_n$
of the chain as $n \sim \sqrt N$. And for L\'evy sums,
it is also well known that the maximal length $l_{max}$ among
the $n$ terms of the sum $N \sim l_1+l_2+ ... l_n$
actually represents a finite fraction
of the sum \cite{Der-levy}, i.e. the biggest loops
indeed occupy a finite fraction of the sample.

With the measure (\ref{critilambdapur}), 
the crossing values
(\ref{Bc},\ref{Rc}) 
are $B(T_c)= 1/3$ and $R(T_c)=9/5$.
For our sampling procedure with $k_{max}$ terms, 
the crossing values are given in eq.(\ref{crossing}), where the 
moments (\ref{moments2}) obtained from (\ref{critilambdapur}) read
 \begin{eqnarray}
\label{puremoments}
<\lambda^m>_{c,k_{max}}= \displaystyle \sum_{k=1}^{k_{max}}
\left( \frac{k}{k_{max}} \right)^{m- 3/2} 
\end{eqnarray}

\begin{figure}[htbp]
\includegraphics[height=6cm]{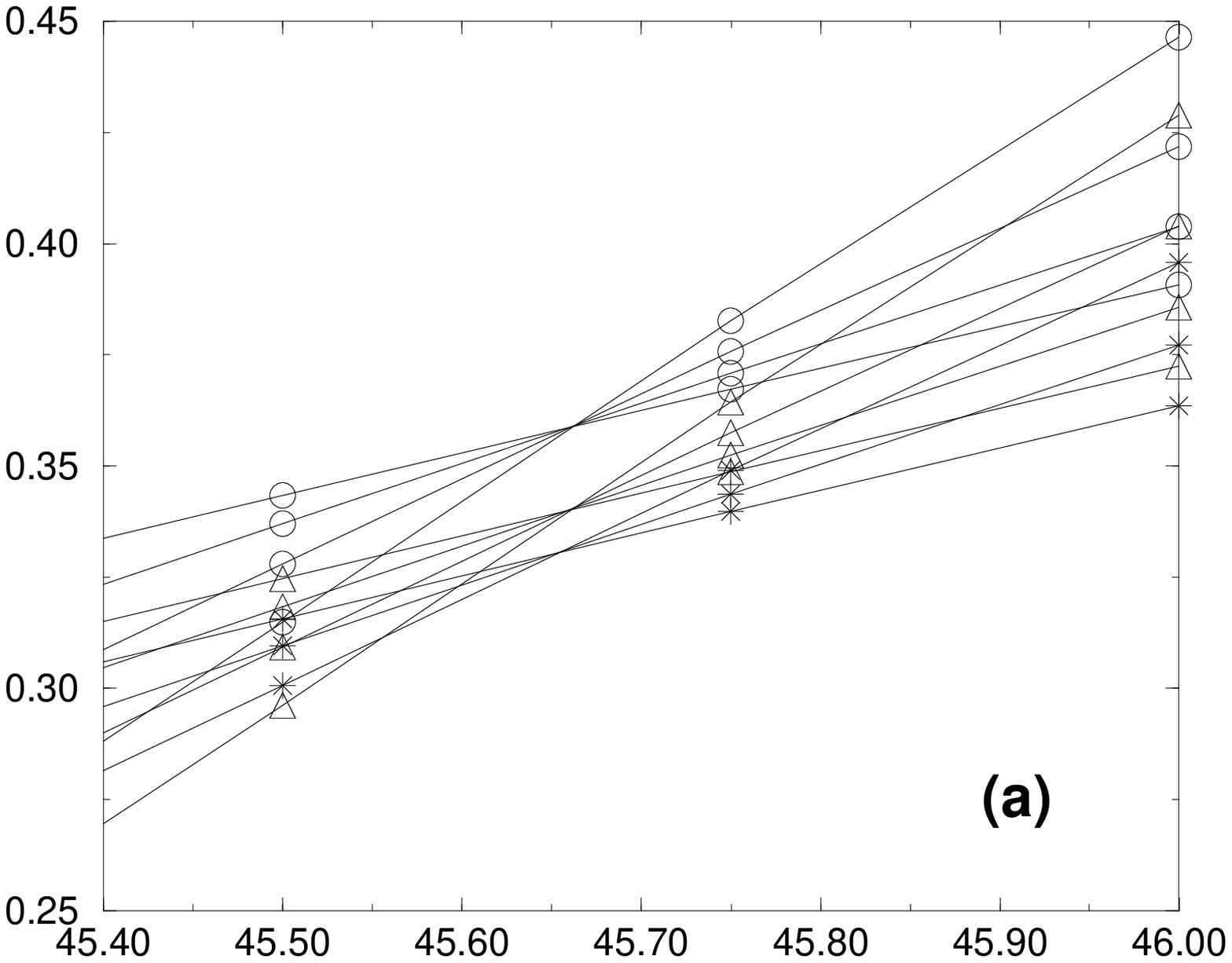}
\hspace{1cm}
\includegraphics[height=6cm]{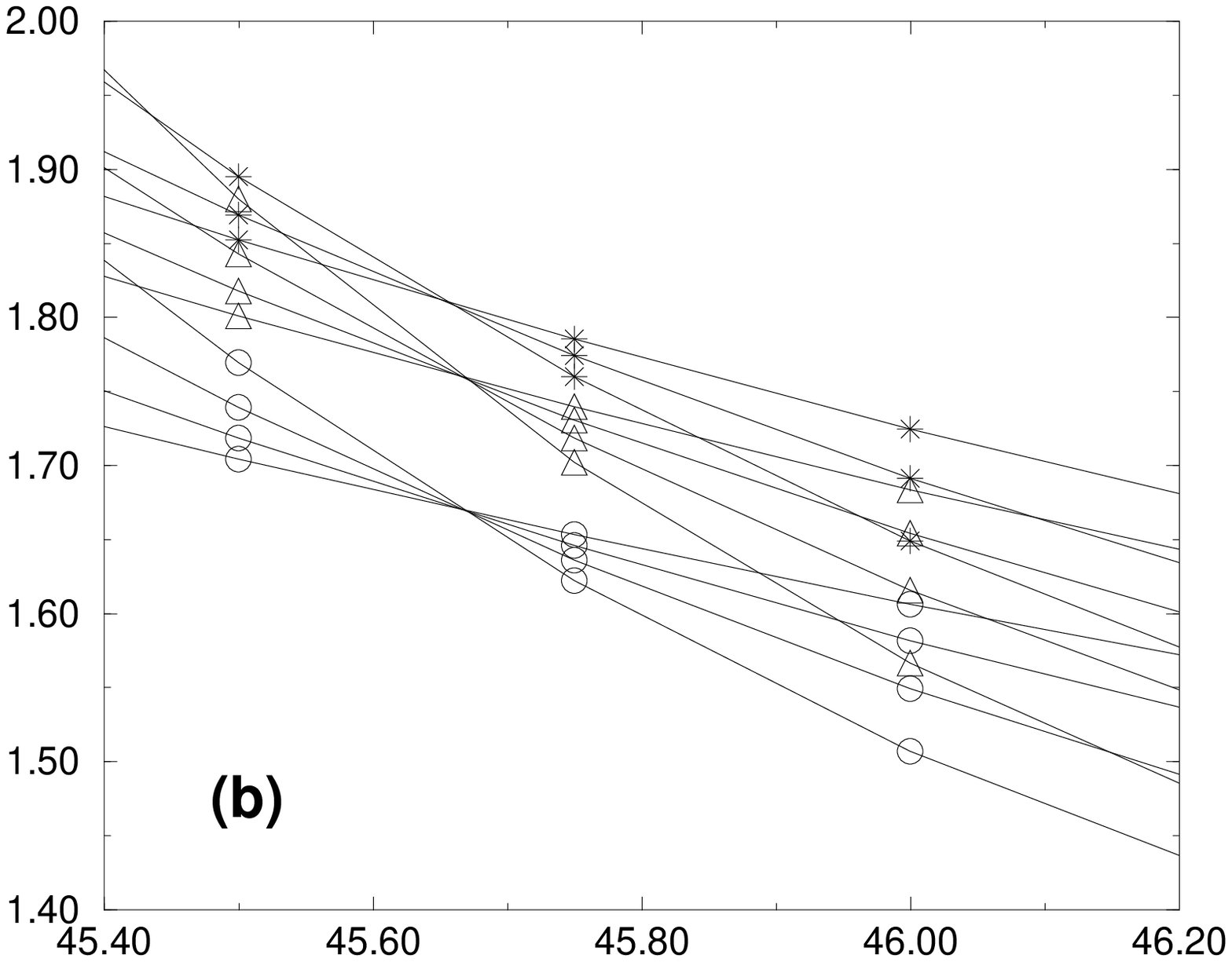}
\caption{(a) The Binder parameter $B_{N}^{(k_{max})}(T)$ of the pure case for
$k_{max}=100$ $(\bigcirc)$, $1000$ $(\bigtriangleup)$, $\frac{N}{2}$
$(\ast)$ and chain lengths  up to $N=8 \cdot 10^5$.
The measured crossing values at the common critical temperature $T_c$
are in agreement with eqs (\ref{crossing},\ref{puremoments}) which
give 0.3612..($k_{max}=100$), 0.3413..($k_{max}=1000$), and $\frac{1}{3}$
($k_{max}=\frac{N}{2}$). 
(b) The Binder parameter $R_{N}^{(k_{max})}(T)$ of the pure case  for
$k_{max}=100$ 
$(\bigcirc)$, $1000$ $(\bigtriangleup)$, $\frac{N}{2}$ $(\ast)$ and
chain lengths up to $N=8 \cdot 10^5$ . The measured crossing values at the common critical temperature $T_c$
are in agreement with eqs (\ref{crossing},\ref{puremoments}) which
give 1.6699..($k_{max}=100$), 1.7584..($k_{max}=1000$), and $\frac{9}{5}$
($k_{max}=\frac{N}{2}$).}
\label{f1}
\end{figure}

We show in Figure \ref{f1} the results of our simulations for the modified
Binder parameters (\ref{defb}) for $k_{max}=100$,
$k_{max}=1000$, and compare them with the calculations of the full
Binder parameters (\ref{defbtrue}), which correspond to
$k_{max}=\frac{N}{2}$. The crossing temperature $T_c$ is indeed
independent of $k_{max}$, and the crossing values are in agreement with
eqs (\ref{crossing},\ref{puremoments}).

Let us now briefly describe the properties of the loop measure
off criticality. The finite-size scaling form of the partition function
$Z_N$ with both ends fixed on the substrate is \cite{loop-caspur} 
 \begin{equation}
Z_N^{pure}(T) \simeq \frac{ 2^N }{ \sqrt N } Q \left( (T-T_c) \sqrt N \right)
\label{fssznpur}
\end{equation}
where the function $Q(x)$ satisfies 

(i) $Q(x=0)=1$ to recover the critical partition function
(\ref{znpurtc}).

(ii) $Q(x \to - \infty)= -x \ e^{x^2} $ that corresponds to
the localized phase
 \begin{equation}
Z_N^{pure}( T<T_c) \opsimeq_{N \gg 1/(T-T_c)^2}  (T_c-T) 2^N e^{ (T-T_c)^2 N } 
\label{znpurloc}
\end{equation}
In this regime, the loop measure  (\ref{pnl}) reads
 \begin{equation}
M_N^{T<T_c}(l) \opsimeq_{N \gg 1/(T-T_c)^2} N (T_c-T)  
\frac{e^{-(T_c-T)^2 l}}{ l^{3/2} }
\label{loclpur}
\end{equation}
i.e. there exists an extensive number $N (T_c-T)$
of finite loops distributed with 
$\rho_{loc}(l) = \frac{ e^{-(T_c-T)^2 l}}{ l^{3/2} }$. The Binder
parameters (\ref{defbtrue}) thus converge as $N \to \infty$ 
towards $B_{ \infty} (T<T_c)=0$ and $R_{ \infty} (T<T_c)=3$.

(iii) $Q(x \to +\infty)=1/x^2$ that corresponds to delocalized phase \cite{Huse,Der_Hak_Van}
 \begin{equation}
Z_N^{pure}(T>T_c) \opsimeq_{N \gg 1/(T-T_c)^2 }
 \frac{ 2^N }{ (T-T_c)^2 N^{3/2} } 
\label{znpurdeloc}
\end{equation}
The factor $1/N^{3/2}$ means that the substrate becomes repulsive
 in the delocalized phase \cite{Huse,Der_Hak_Van}.
In this regime, the loop measure becomes concentrated on $\delta(\lambda-1)$
and the associated Binder parameters (\ref{defbtrue}) converge towards
$B_{N \to \infty} (T>T_c)=1$ and $R_{N \to \infty} (T>T_c)=1$.

\begin{figure}[htbp]
\includegraphics[height=6cm]{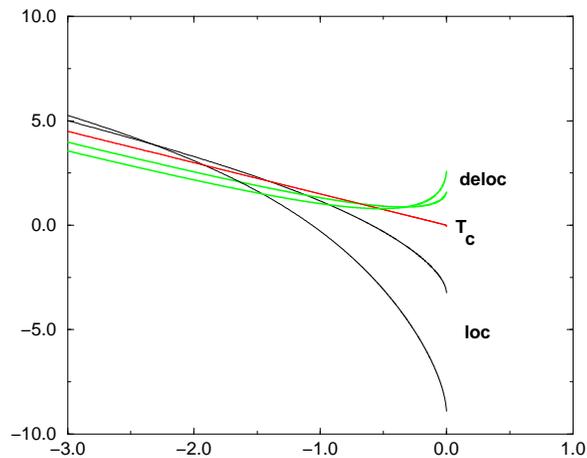}
\caption{Log-Log plot of the loop measure $M_N^{T}(\lambda)$ of the
pure case for
$N=10^5$ and $N=4 \ \cdot 10^5$, and $T=T_c-2$ (loc), $ T_c$ and
$T_c+2$ (deloc). At $T_c$, the two curves are superimposed. Above
$T_c$, a minimum appears.}
\label{separatricecaspur}
\end{figure}

We show on Figure \ref{separatricecaspur} the loop measures
 ${\cal M}_N^T(\lambda)$ for two sizes ($N=10^5$ and $N=4 \ \cdot
10^5$), below, at and above $T_c$. At $T_c$, the loop measure is
independent of $N$ and corresponds to eq (\ref{critilambdapur}). Above
$T_c$, an $N$-dependent minimum shows up.

\subsection{ Disordered case}

We have numerically studied the wetting transition
for the following binary distribution for the contact
energies ($\varepsilon_{\alpha}$)
 \begin{equation}
P (\varepsilon_{\alpha})= (1-p) \delta(\varepsilon_{\alpha}-\varepsilon_{0})
+ p \delta(\varepsilon_{\alpha})
\label{defdilution}
\end{equation}
with the three dilution fractions $p=0.25$, $p=0.5$ and $p=0.75$.

 We have used the sampling
method explained above for the loop statistics
with the factor $k_{max}=1000$. We have
computed the modified Binder parameters $B_{N}^{(k_{max})}(T)$ and
$R_{N}^{(k_{max})}(T)$ (\ref{defb}), for each disorder 
sample, and for sizes $N=10^5, \  2 \cdot 10^5, \ 4 \cdot 10^5$. Both
quantities have been then averaged over $10^4$ independent samples 
(from now on, $\overline{A}$ denotes the disorder average of the
quantity $A$). The crossing of $\overline{B_{N}^{(k_{max})}}(T)$ and
$\overline{R_{N}^{(k_{max})}}(T)$ then yield reasonable error bars in the
localization of $T_c$ (see Figure \ref{f3}).

\subsubsection{Self-averaging properties}

\begin{figure}[htbp]
\includegraphics[height=6cm]{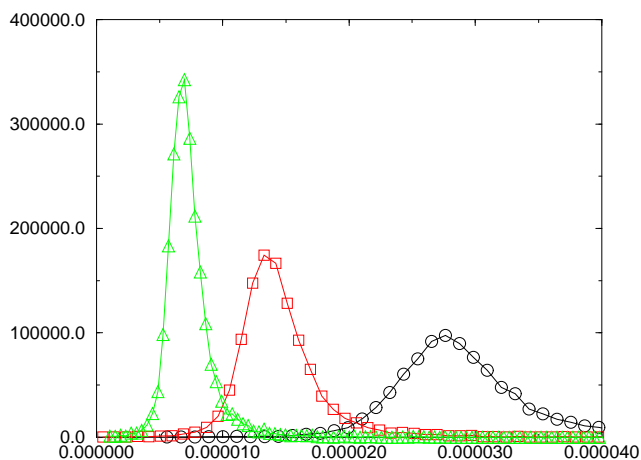}
\caption{ Histogram over $10^4$ disordered samples of $M_N(l=\frac
{N}{2})$ at criticality in the case $p=0.5$, for
sizes $N=10^5$ $(\bigcirc)$, $2 \cdot 10^5$ $(\square)$, $4 \cdot 10^5$ 
$(\triangle)$.}
\label{f11}
\end{figure}

In disordered systems, extensive quantities are expected to be
self-averaging, because spatial averages in a given sample are
equivalent in the thermodynamic limit to disorder averages. For
correlation functions, the situation is more subtle, as 
discussed in details in \cite{Der_Hil} for spin-spin correlations in magnetic systems.
Since the loop measure $M_N(l)$ (\ref{pnl})
is a spatially averaged multiple point correlation function, 
we have studied its self averaging properties. We show in Figure
\ref{f11} the histogram over $10^4$ disordered samples of $M_N(l=\frac
{N}{2})$ at criticality in the case $p=0.5$, for sizes $N=10^5, \ 2
\cdot 10^5, \ 4 \cdot 10^5$: the distribution
of $M_N(l=\frac{N}{2})$ over the samples is more and more peaked
around its average as $N$ grows.  

Concerning the crossing of Binder parameters, we have also checked
that averaging separately the moments in the numerator and denominator
in eqs. (\ref{defbtrue}) gives the same values as the averaged Binder
parameters, e.g.
\begin{eqnarray}
\overline{B_N}(T) \simeq  \frac{\overline{<l^2>}}{\overline{<l>} N} 
\label{defbtrue2}
\end{eqnarray} 
This property was also found for usual Binder parameters in magnetic
systems \cite{Rie_You}.

\subsubsection{Binder parameters crossings}

\begin{figure}[htbp]
\includegraphics[height=6cm]{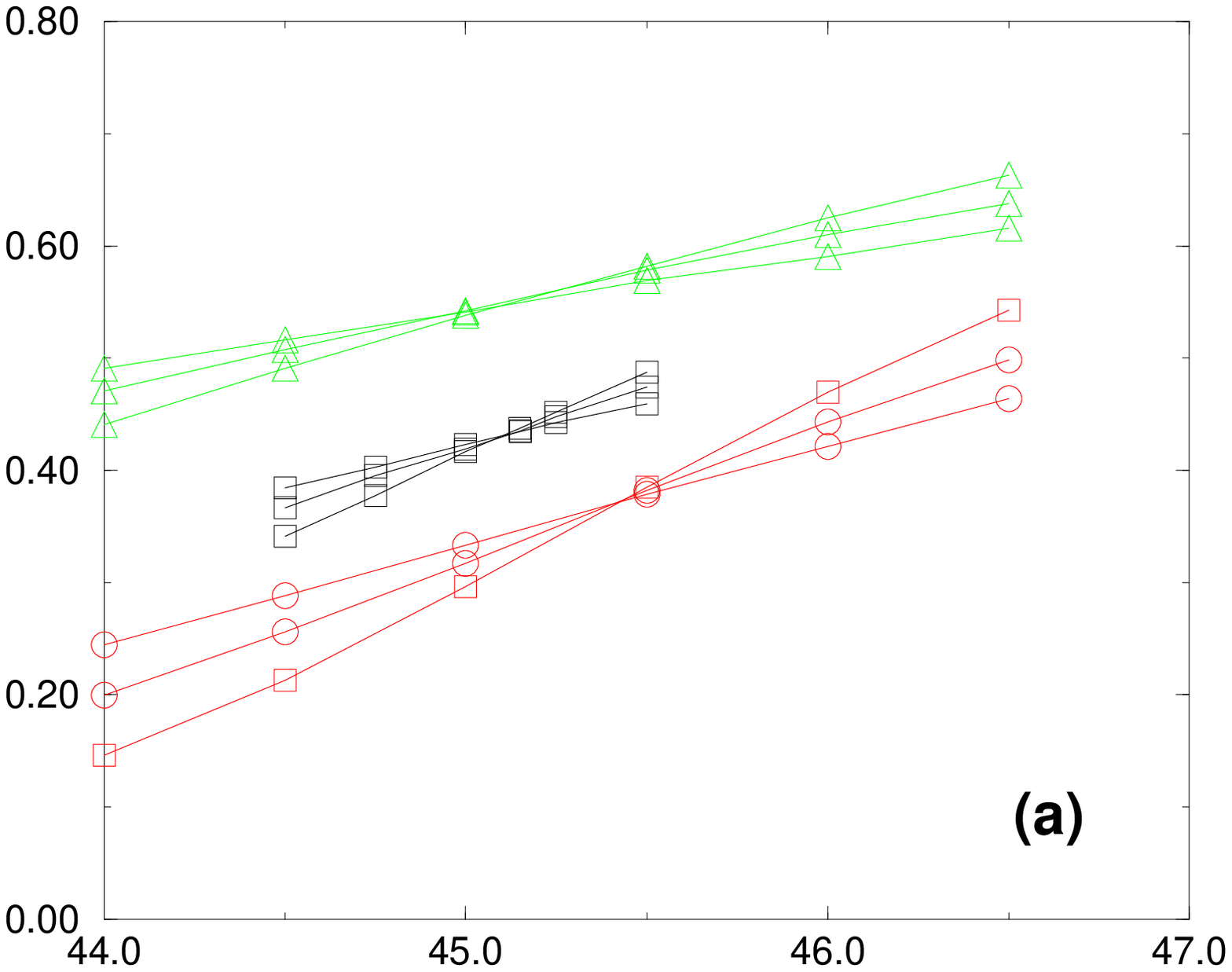}
\hspace{1cm}
\includegraphics[height=6cm]{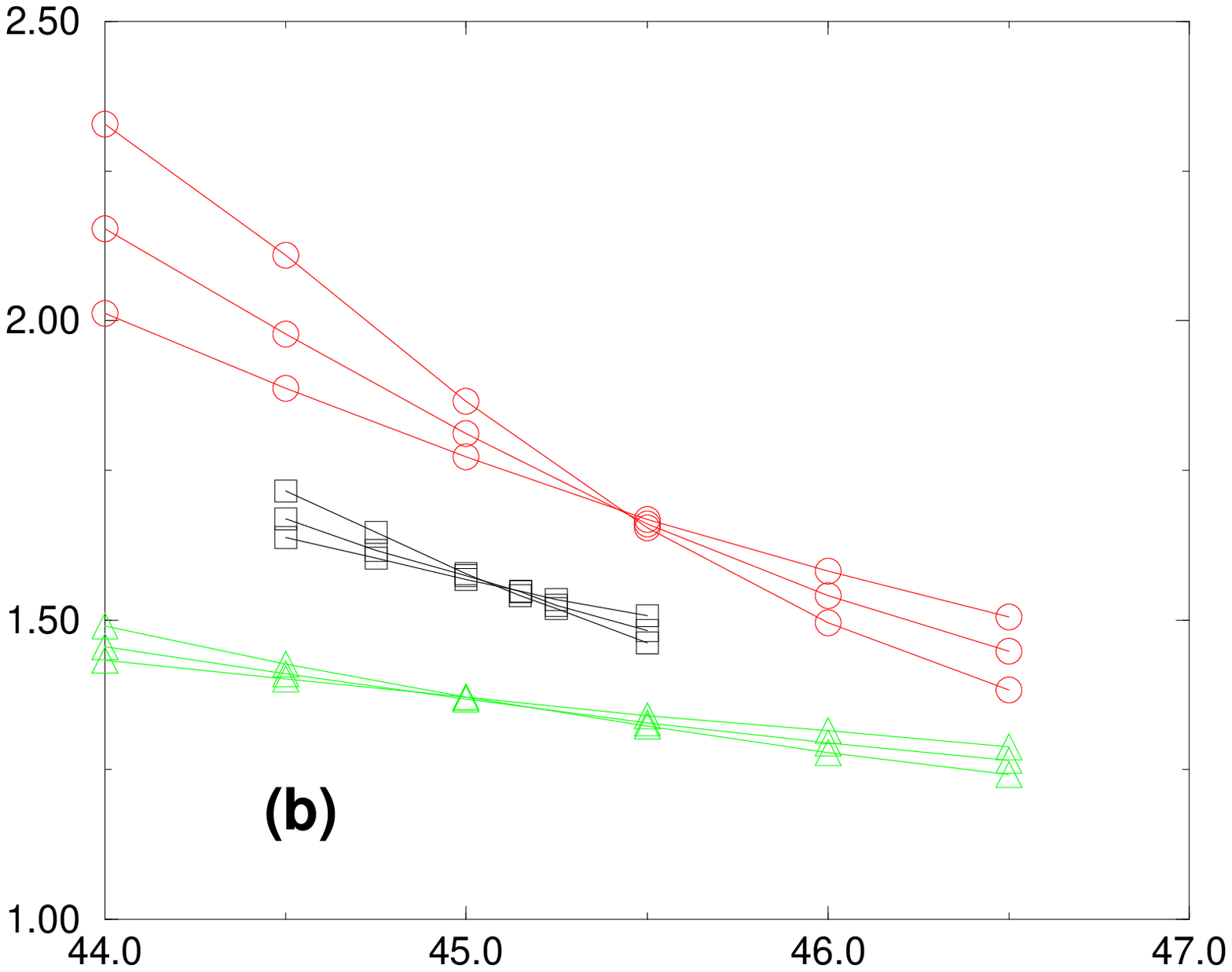}
\caption{(a) The averaged Binder parameter
$\overline{B^{(k_{max})}}_N(T)$, for $k_{max}=1000$, ($N=10^5$, $2
\cdot 10^5$, $4 \cdot 10^5$), and $p=0.25 \ (\bigcirc), \ 0.5 \
(\square), \ 0.75 \ (\triangle)$. We 
have rescaled 
the contact energy $\varepsilon_0$ so that the different $Tc(p)$ are
close. The error bars are much smaller than the symbols.
(b) The averaged Binder parameter $\overline{R^{(k_{max})}}_N(T)$, for
$k_{max}=1000$, ($N=10^5$, $2
\cdot 10^5$, $4 \cdot 10^5$), and $p=0.25 \ (\bigcirc), \ 0.5 \
(\square), \ 0.75 \ (\triangle)$.}  
\label{f2}
\end{figure}

\begin{figure}[htbp]
\includegraphics[height=6cm]{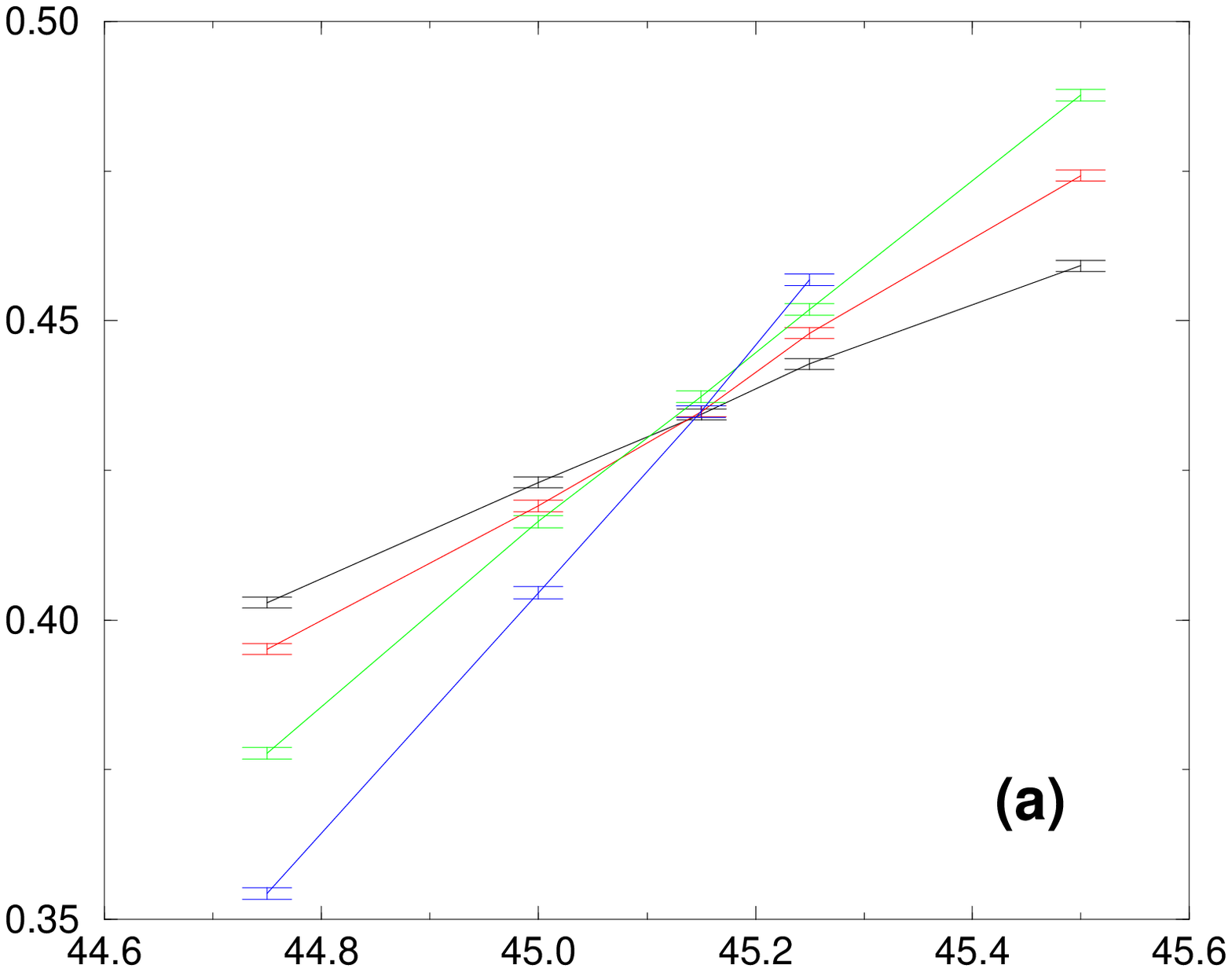}
\hspace{1cm}
\includegraphics[height=6cm]{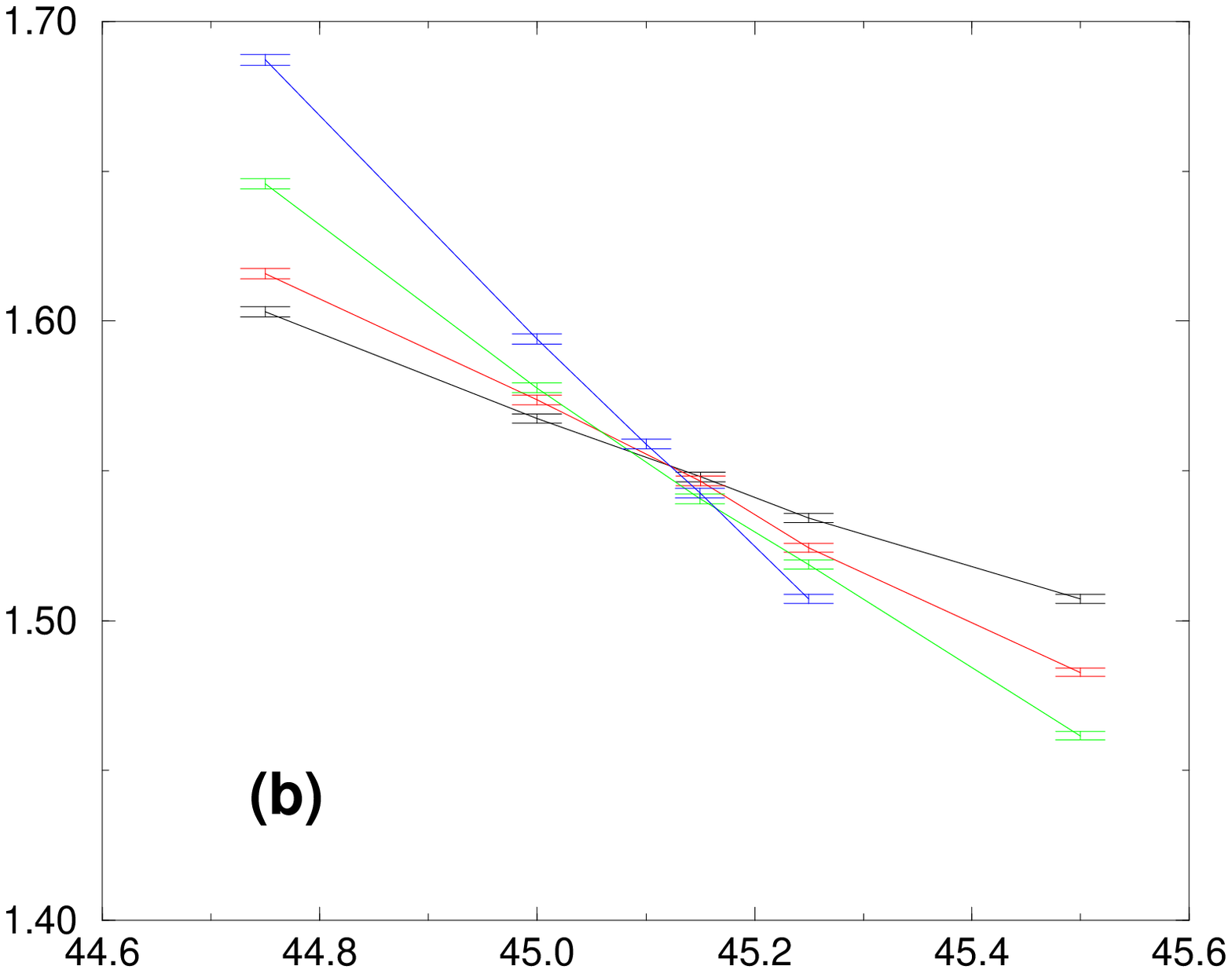}
\caption{(a)  The averaged Binder parameter $\overline{B^{(k_{max})}_N}(T)$
, for $k_{max}=1000$ and $N=1, 2, 4, 8 \cdot
10^5$ and $p=0.5$ (b) The averaged Binder parameter
$\overline{R^{(k_{max})}_N}(T)$, for $k_{max}=1000$ and $N=1, 2, 4, 8 \cdot
10^5$ and $p=0.5$.} 
\label{f3}
\end{figure}

In Figure \ref{f2}, we show the results for the averaged Binder parameters
$\overline{B_{N}^{(k_{max})}}(T)$ and 
$\overline{R_{N}^{(k_{max})}}(T)$ for sizes $N =1, 2, 4 \cdot
10^5$. The numerical values at the crossing depend on 
the dilution fraction $p$ (\ref{defdilution}) :

(i) for $p=0.25$, we obtain $\overline{B^{(k_{max})}}(T_c)=0.370 \pm 0.005$ and
$\overline{R^{(k_{max})}}(T_c)=1.685 \pm 0.005$.
The critical temperature $T_c=45.43 \pm 0.03$ 
is slightly below the annealed temperature $T_{ann} =45.50$ (here
$\varepsilon_0=-270$).

(ii) for $p=0.5$, where we have also studied $N=8 \cdot 10^5$,  
we obtain $\overline{B^{(k_{max})}}(T_c)=0.430 \pm 0.005$ and
$\overline{R^{(k_{max})}}(T_c)=1.547 \pm 0.005$, 
with $T_c=45.13 \pm
0.03$ as compared to the annealed temperature $T_{ann} =45.42$.
(here $\varepsilon_0=-350$).

(iii) for $p=0.75$, we obtain $\overline{B_{N}^{(k_{max})}}(T_c)=0.550 \pm
0.005$ and $\overline{R_{N}^{(k_{max})}}(T_c)=1.360 \pm 0.005$, 
with $T_c=45.15 \pm 0.05$ as compared to the annealed temperature
$T_{ann} =46.82$ (here $\varepsilon_0=-515$). 

These results show that the crossings of Binder parameters
allows to locate precisely the critical temperature.
We now turn to the analysis of critical properties of various observables.

\section{Study of critical properties}

\subsection{Distribution of loops of length $l\sim O(N)$ at criticality}

\begin{figure}[htbp]
\includegraphics[height=6cm]{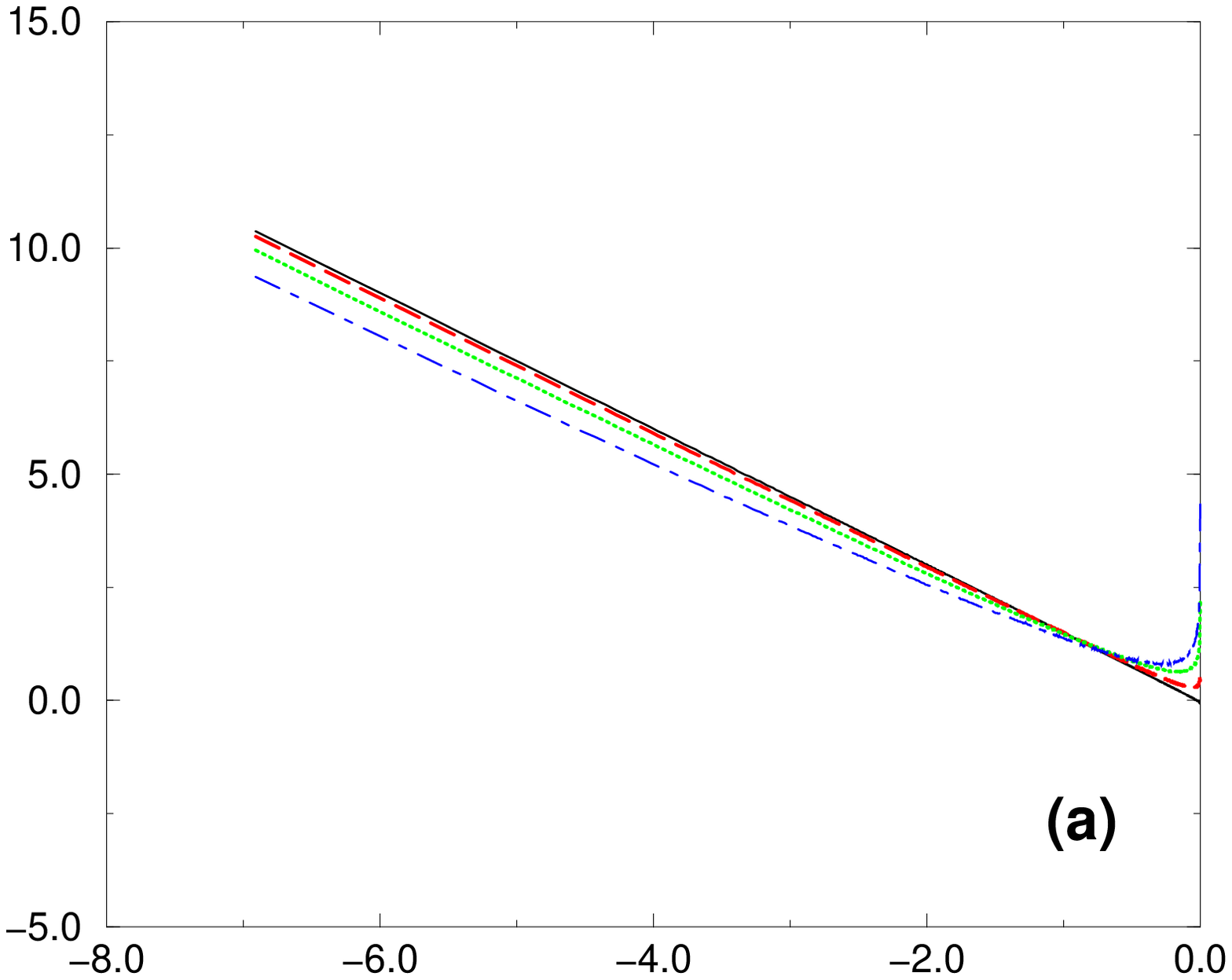}
\hspace{1cm}
\includegraphics[height=6cm]{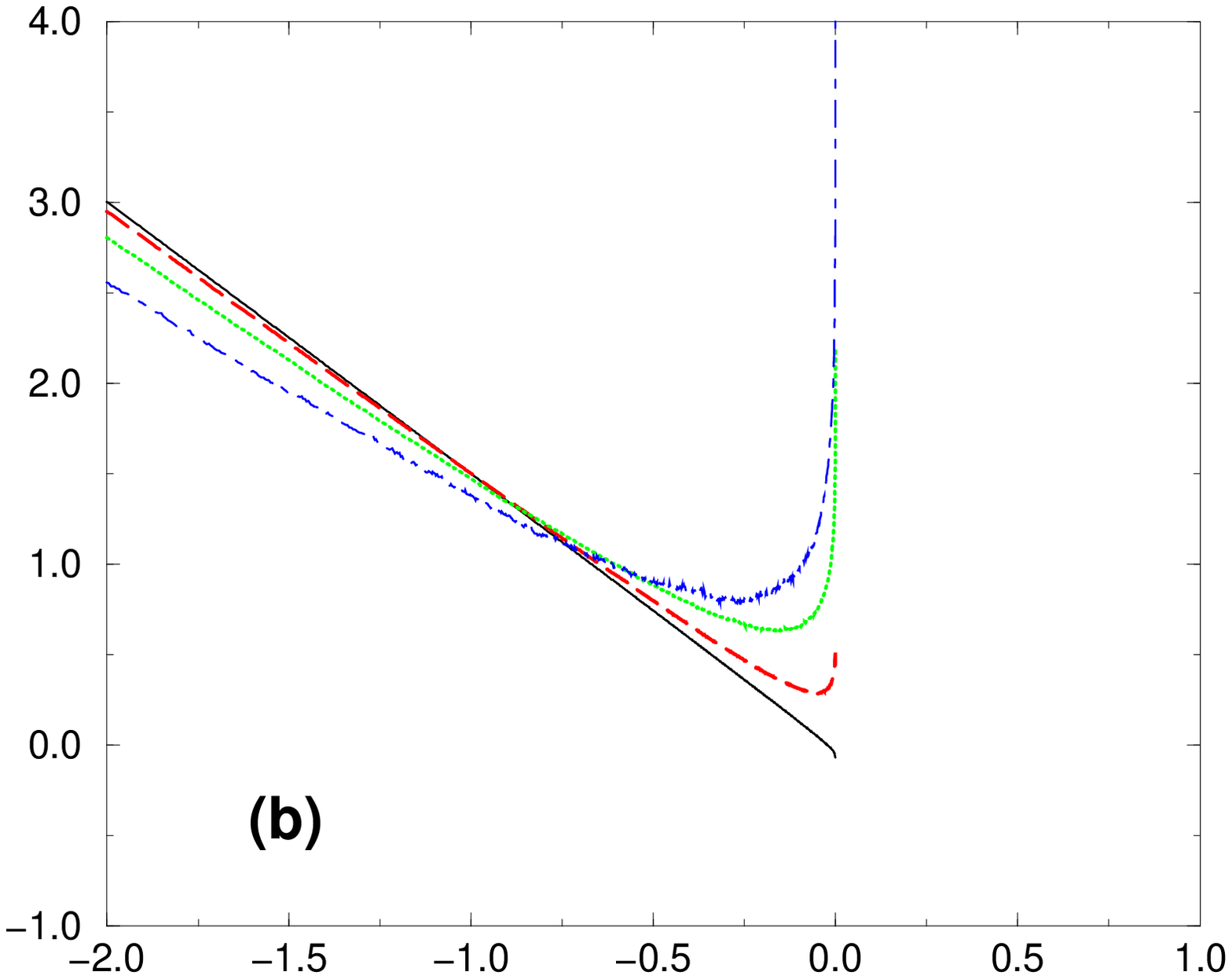}
\caption{(a) Log-Log plot of $\overline{\cal M}_c(\lambda)$ for $p=0.25,
\ 0.5 , \ 0.75$. The pure case $p=0$ (solid black line) is shown for
comparison (b) Zoom near $\lambda \to 1$.}
\label{f4}
\end{figure}

The crossings values $(B_c,R_c)$ of the Binder parameters
vary continuously with the dilution fraction $p$ (Figure \ref{f2}).
This means that the measure $\overline{{\cal M}_c}(\lambda)$ of loops
occupying a finite fraction $\lambda=l/N$ of the sample at criticality
also depends continuously on $p$.
We present on Figure \ref{f4}
these loop measures $\overline{{\cal M}_c}(\lambda)$
for $p=0.25$, $p=0.5$ and $p=0.75$ in log-log plot, and compare them
with the pure case $p=0$, which corresponds to
the straight line $(\ln {\cal M}_c(\lambda) = -(3/2) \ln \lambda$, eq
(\ref{critilambdapur})).   
In the limit $\lambda \to 0$ (or $\ln \lambda \to -\infty$),
the measures $\overline{{\cal M}_c}(\lambda)$ become asymptotically parallel
to the pure case for all $p$, i.e.
\begin{eqnarray}
\overline{{\cal M}_c} (\lambda) \oppropto_{\lambda \to 0} \frac{1}{
\lambda^{3/2} }
\label{pltc33}
\end{eqnarray} 
(see also the more detailed study of finite loops below),
but otherwise, the loop measures in the disordered cases are
qualitatively different from the pure case: a minimum occurs, followed
by a weak divergence as $\lambda \to 1$ (Figure \ref{f4} (b)). 
The form of the divergence suggests that it is logarithmic
with a $p$ dependent exponent.
The simplest form that can represent the critical loop measure
$\overline{{\cal M}_c}(\lambda)$ on the full interval $0<\lambda<1$
and that is compatible with all our data is
\begin{eqnarray}
\overline{ {\cal M}_c} (\lambda) \simeq \frac{1}{{\lambda}^{3/2}}
\left(1+
\frac{C(p)}{(-\ln \lambda)^{\delta(p)}}\right)  
\label{pltc12}
\end{eqnarray} 
The values $\delta(p=0.25) \simeq 0.1$, $\delta(p=0.5) \simeq 0.25$ and
$\delta(p=0.75) \simeq 0.4$ for the exponent $\delta$, and the common
value $C(p=0.25) \sim C(p=0.5) \sim C(p=0.75) \sim 6$  
for the amplitude $C$ give good fits of (i) the measures $\overline{ {\cal
M}_c} (\lambda)$ on the whole range $\lambda \in [0,1]$, with a
correct location of the $p$ dependent minimum, and (ii) to the values of the
Binder parameters crossings shown in Figures \ref{f2} and \ref{f3}. 

\begin{figure}[htbp]
\includegraphics[height=6cm]{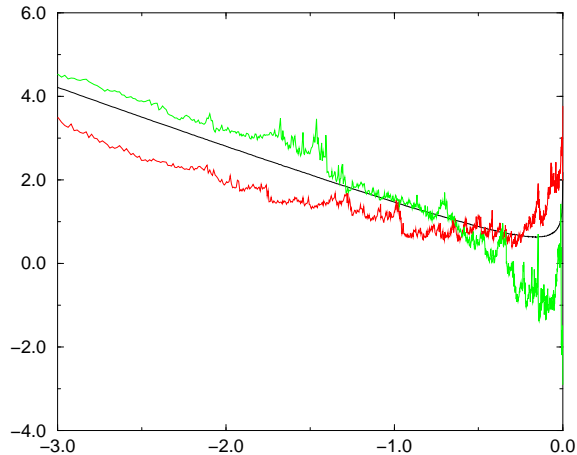}
\caption{ At criticality for the case $p=0.5$:
 Log-Log plot of the averaged loop measure $\overline{{\cal
M}_c}(\lambda)$ over $10^4$ samples, as compared to the loop measure
of two particular samples. The upward (resp. downward) oriented curve
points towards a delocalized (resp. localized) trend}
\label{fluct}
\end{figure}

Beyond this numerical evidence,
it would be of course  interesting to have a theoretical explanation
for the appearance of this logarithmic singularity 
in the loop measure near $\lambda \to 1$ in the disordered case.
The only qualitative argument we can think of at this stage is the following :
in the pure case, we have seen that a minimum appears
in the loop measure in the {\it delocalized} phase
(see Figure \ref{separatricecaspur}).
In the disordered case, one may argue that the minimum
of the disordered averaged loop measure $\overline{ {\cal M}_c} (\lambda)$
at criticality comes from the fact that at $T_c$, among the disordered
samples of size $N$, a fraction of these samples tend to be slightly
delocalized, with a minimum in their loop measure $ {\cal M}_c (\lambda)$,
whereas the other samples tend to be slightly localized
(see Figure \ref{fluct}).
In other words, if one imagines to associate to each sample $i$
a sample-dependent pseudo critical temperature $T_c^N(i)$, as it was
done in other disordered systems \cite{domany}, the presence of the
minimum at $\lambda_{min}<1$ reflects the spreading of the pseudo
critical temperatures $T_c^N(i)$ around the thermodynamic critical
temperature $T_c$. The fact that the exponent $\delta(p)$ grows with
$p$ could be interpreted as a consequence of a growing dispersion of
the pseudo-critical temperatures $T_c^N(i)$ with the strength of the
disorder. 

\subsection{Measure of finite loops at criticality}

In the section above, we have discussed the statistics
of loops of length $l \sim O(N)$.
We now discuss the statistical properties of
 finite loops, i.e. of length $1 \ll l \ll N$. 
We have measured the
critical loop measure $M_N(l)$ for $N= \ 1, \ 2,\ 4 \ \cdot 10^5$ with
fixed values of $l$ ( $l=10 \ k \ ; \ k=1,2,...,1000$).
We obtain that the dependence in $l$ is a pure power law
\begin{equation}
\label{fixedtc}
\overline{M_N^{Tc}}(l) \propto \frac{1}{{l}^{\gamma(N)}}
\end{equation}
with an effective exponent $\gamma(N)$ which increases towards the pure
value $3/2$ as $N$ increases. For instance for $p=0.5$, we get
$\gamma(N=10^5) \simeq 1.46$, $\gamma(N=2 \cdot 10^5) \simeq 1.468$,
$\gamma(N=4 \cdot 10^5) \simeq 1.475$. 
This power-law behavior with exponent $3/2$ is moreover 
in agreement with the asymptotic behavior (\ref{pltc33})
for loops of length $O(N)$ in the limit $\lambda \to 0$,
as it should, since
the large $l$ behavior of the finite loop measure $\overline{M_N^{Tc}}(l)$
should match the small $\lambda \to 0$ behavior
of $O(N)$ loop measure $\overline{{\cal M}_c}(\lambda)$.
This requirement actually determines the $N$-normalization
of finite loops.
Indeed, we have obtained that at criticality,
the measure $\overline{{\cal M}_c}(\lambda)$ of $O(N)$ loops
is independent of the size $N$ and is well described 
by the form (\ref{pltc12}). 
Via the change of variable $l=\lambda N$, this leads to the following
normalization for the loop critical distribution
(\ref{fixedtc})
\begin{eqnarray}
\overline{M^{Tc}_N}(l) \opsimeq_{1 \ll l \ll N}
 \frac{\sqrt N}{l^{3/2}} \left(1+
\frac{ C(p) }{(\ln N )^{\delta(p)}}\right)  
\label{mlcriti}
\end{eqnarray}

\subsection{Contact density at criticality}

As explained at the beginning, the contact density
(\ref{theta}) is directly related to the normalization
of the loop measure via (\ref{thetapl}).
The result (\ref{mlcriti}) for the normalization in $N$
of the loop measure thus yields 
the following finite-size behavior
for the contact density at criticality 
\begin{eqnarray}
\label{normalisation}
\overline{\theta_N} (T_c) \oppropto
 \frac{1}{\sqrt N} \left(1+
\frac{ C(p) }{(\ln N )^{\delta(p)}}\right)  
\end{eqnarray} 
i.e. the leading scaling behavior is the same as in the pure case,
but there are strong logarithmic corrections to scaling. Plotting
$({\sqrt N} \ \ 
\overline{\theta_N} (T))$ for various $N$ thus yields a very poor
determination of $T_c$, in marked contrast with its precise location
through the crossings of the Binder parameters, where these
logarithmic corrections are absent.

We have directly computed $\overline{\theta_N}(T_c)$ for $N=10^5$, $2
\cdot 10^5$ and $4 \cdot 10^5$, for $p=0.25, \ 0.5$ and $0.75$. Our
results are in agreement with eq. (\ref{normalisation}), with the same 
values of $C(p)$ and $\delta(p)$ quoted above, just after eq.
(\ref{pltc12}). This further supports
the form (\ref{pltc12}) of the critical loop distribution.

\subsection{Energy at criticality}

We now consider the contact energy (\ref{energy}). In the binary case, 
it is closely related to the contact density (\ref{theta}), since
$\theta_N=\frac{e_N}{\varepsilon_0}+r_N$, where $r_N$ is the contact
density of diluted sites. At criticality, we expect that $r_N$ scales
at most like $e_N$, which implies that the energy $e_N(T_c)$ has the
same finite size properties as $\theta_N(T_c)$ , eq. (\ref{normalisation})
\begin{eqnarray}
\label{energietc}
\overline{e_N} (T_c) \oppropto
 \frac{1}{\sqrt N} \left(1+
\frac{ C_e(p) }{(\ln N )^{\delta(p)}}\right)  
\end{eqnarray} 
with a coefficient $C_e(p) \le C(p)$ .The
direct measure of the ratios $\frac {e_N(T_c)}{\theta_N(T_c)}$
increases very slowly with $N$, and
typical values for $N=2.10^5$ are $0.87$ ($p=0.25$), $0.74$ ($p=0.5$)
and $0.6$ ($p=0.75$).

\subsection{Finite-size scaling in the critical region }

\begin{figure}[htbp]
\includegraphics[height=6cm]{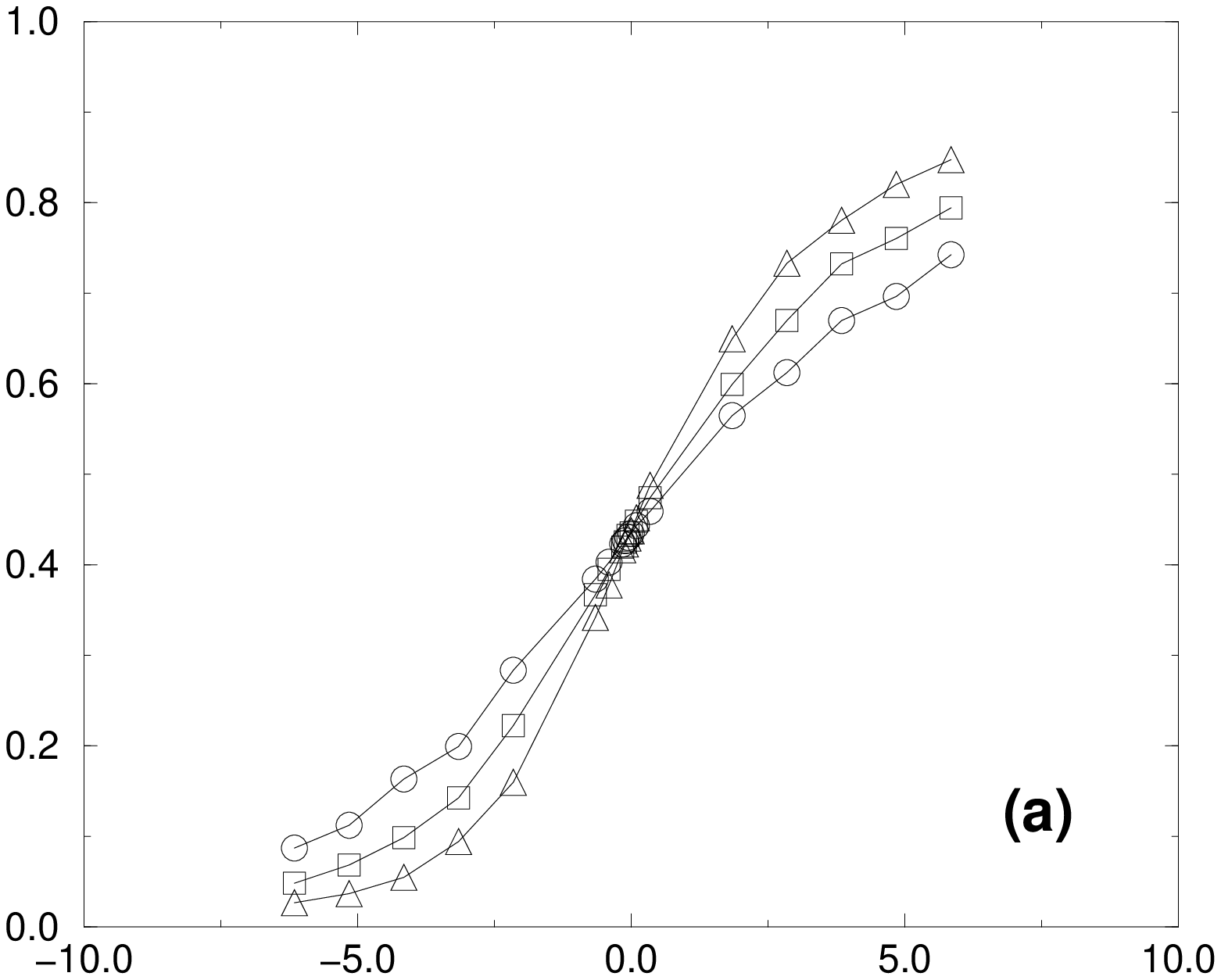}
\hspace{1cm}
\includegraphics[height=6cm]{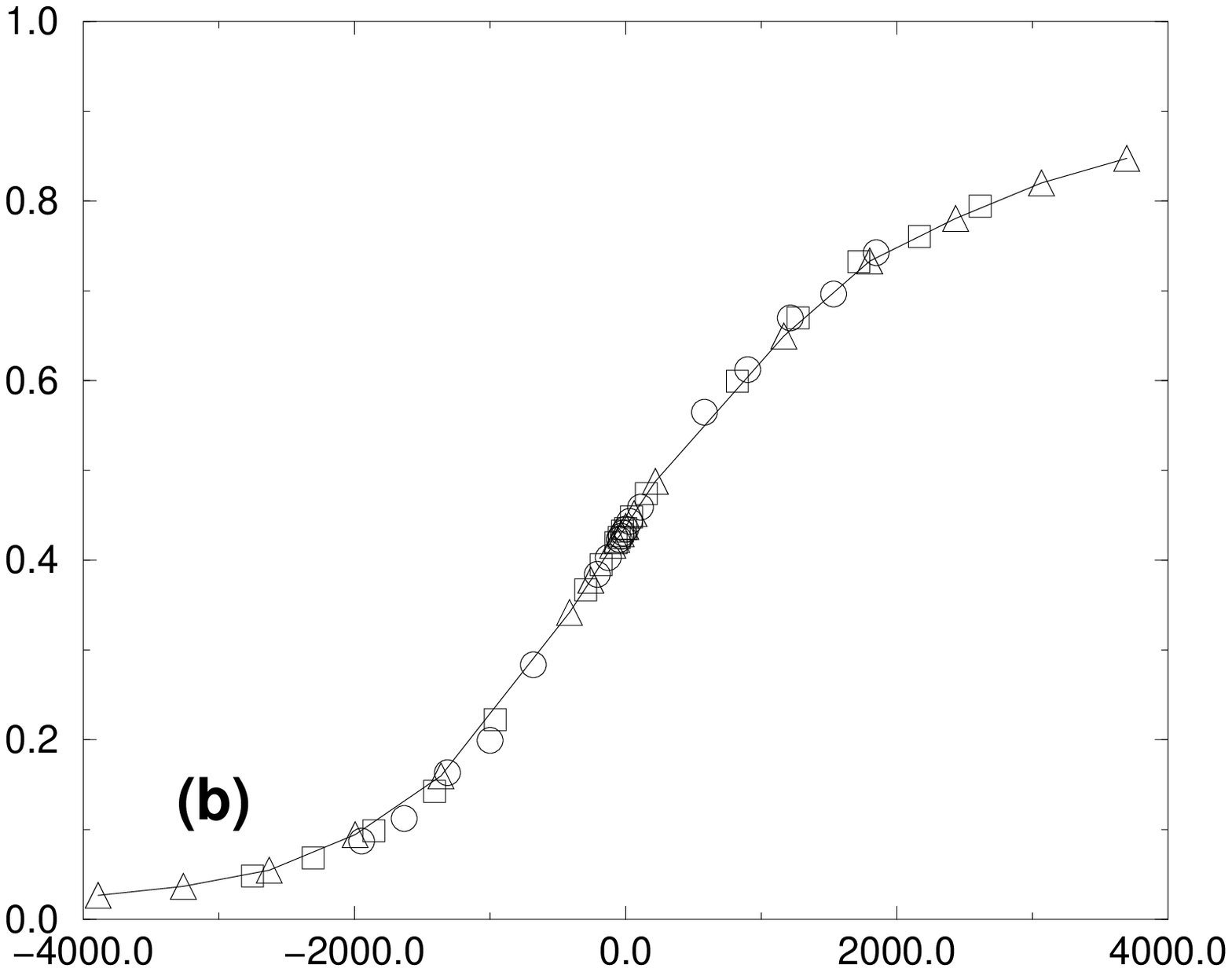}
\caption{(a) The Binder parameter $\overline{B_N}(T-T_c)$ of the
$p=0.5$ case, for $N=10^5$
$(\bigcirc)$, $2 \cdot 10^5$ $(\square)$, $4 \cdot 10^5$  
$(\triangle)$ (b) Master curve ${\cal B} \left( x= (T-T_c) \sqrt{N}
\right)$ of equation (\ref{fssb}) for the same data and symbols.}
\label{master}
\end{figure}

We are now interested into the finite-size scaling in the critical
region. In terms of the specific heat exponent  $\alpha$,
the singularity of the free-energy is $f(T_c)-f(T) \sim
(T_c-T)^{2-\alpha}$. Via hyperscaling ( $f(T) \sim 1/ \xi(T)$ in
dimension $d=1$), the correlation length along the interface diverges as
$\xi(T) \sim 1/(T_c-T)^{2-\alpha}$.
According to finite-size scaling theory,
the appropriate rescaled variable is the ratio $N/\xi(T)$
between the size $N$ of the system and this correlation length $\xi(T)$.
As a consequence, we expect that the Binder parameter
$\overline{ B_N(T) } $ obtained for various sizes $N$
actually only depend on the ratio $N/\xi(T) \sim N (T_c-T)^{2-\alpha}$ or equivalently
 \begin{eqnarray}
\overline{B_N}(T)= {\cal B} \left( x= (T-T_c) N^{ \phi} \right)
 \ \   {\rm with } \ \   \phi= \frac{1}{2 -\alpha}
\label{fssb}
\end{eqnarray}

We show on Figure \ref{master} the master curve obtained with
crossover exponent $\phi=1/2$, corresponding to $\alpha=0$.

Considering now the finite size scaling for the energy, equations
(\ref{energietc}) for the energy at $T_c$ and the crossover exponent
found in (\ref{fssb}) suggest the following form
\begin{equation}
e_N(T)=\frac{1}{\sqrt N} \left[ G_0((T_c-T) \sqrt {N}) +\frac{1}{(\ln
N)^{\delta(p)}}G_1((T_c-T) \sqrt {N}) +.....\right]
\end{equation}
where the $...$ represent higher order terms. Our conclusion is that
the critical exponents of the binary disordered case are the same as
those of the pure case, except for strong corrections to scaling that
are $p$ dependent. 

\section{Conclusions and perspectives}

We have studied the two dimensional wetting transition, in the
presence of binary disorder for various dilution fraction $p$. Our
analysis is based on the probability measure for the loops of length
$l$ existing in a sample of size $N$, with both ends of the chain fixed on
the substrate. We have first shown how the introduction of Binder-like
parameters, built out of the first moments of the loop 
measure, allows to locate precisely the critical temperature. We have
then found numerical evidence that the critical loop distribution
$\overline{{\cal M}_c}(\lambda)$ in the rescaled variable $\lambda=l/N
\in [0,1]$ is not a pure power law (in contrast with the pure
case), but contains a logarithmic divergence near $\lambda \to 1$, with a 
$p$ dependent exponent $\delta(p)$. Finally, we have explained how
this singularity in the loop measure induces very strong logarithmic
corrections to scaling for the contact density, for the energy, and
more generally for thermodynamic quantities. 

Our present analysis of the binary disordered case raises the question
of the dependence of the critical behavior on the disorder
distribution. Indeed, we have obtained that the
critical loop distribution $\overline{{\cal M}_c}(\lambda)$ varies
continuously with the dilution fraction $p$ of the binary distribution.
More generally, we might expect that $\overline{{\cal M}_c}(\lambda)$
could depend upon the disorder distribution itself. If this is the 
case, do the critical exponents differ from those of the binary case?
Since the example of Gaussian disorder \cite{Ta_Cha} has been
interpreted in terms of essential singularities of the
Kosterlitz-Thouless type, we intend to study other types of disorder
distributions to clarify this issue.

Another interesting direction concerns the effect of
disorder when the a priori loop entropy (\ref{asymp}) in the
Poland-Scheraga formulation has an exponent $c>2$,
in which case the pure transition is first-order \cite{Pol_Scher}.
Indeed, in the context of the DNA denaturation transition,
the binding transition between two pure self-avoiding chains
on a cubic lattice was found to be first order \cite{Barbara_pure}.
The theoretical explanation that has been proposed  \cite{Ka_Mu_Pe},
is that the self-avoidance constraint between denaturated loops and
the rest of the chain actually induces an exponent $c>2$ for the loop
weight (\ref{asymp}). The value  $c \sim 2.11$ has been since measured
in Monte-Carlo simulations \cite{Orlandini}. In the future, we hope to
apply our method to the disordered Poland-Scheraga model for the case
$c>2$, and to compare with the results recently obtained by B. Coluzzi
\cite{Barbara} via Monte-Carlo simulations of self avoiding walks.

{ \bf Acknowledgements : } It is a pleasure to thank B. Coluzzi
and J. Houdayer for useful discussions, as well as
H. Orland for many contributions over the years.

\end{document}